\documentclass[aps,pra,twocolumn,groupedaddress,showpacs]{revtex4}
\usepackage{amsfonts,amssymb,amscd,amsmath}
\usepackage{graphicx}
\usepackage{epsfig}
\usepackage{subfigure}
\usepackage{latexsym}
\usepackage{amssymb}
\usepackage{setspace}
\usepackage{color}

\begin{document}

\def\abs#1{ \left| #1 \right| }
\def\lg#1{ | #1 \rangle }
\def\rg#1{ \langle #1 | }
\def\lrg#1#2#3{ \langle #1 | #2 | #3 \rangle }
\def\lr#1#2{ \langle #1 | #2 \rangle }
\def\me#1{ \langle #1 \rangle }

\newcommand{\bra}[1]{\left\langle #1 \right\vert}
\newcommand{\ket}[1]{\left\vert #1 \right\rangle}
\newcommand{\bx}{\begin{matrix}}
\newcommand{\ex}{\end{matrix}}
\newcommand{\be}{\begin{eqnarray}}
\newcommand{\ee}{\end{eqnarray}}
\newcommand{\nn}{\nonumber \\}
\newcommand{\no}{\nonumber}
\newcommand{\de}{\delta}
\newcommand{\lt}{\left\{}
\newcommand{\rt}{\right\}}
\newcommand{\lx}{\left(}
\newcommand{\rx}{\right)}
\newcommand{\lz}{\left[}
\newcommand{\rz}{\right]}
\newcommand{\inx}{\int d^4 x}
\newcommand{\pu}{\partial_{\mu}}
\newcommand{\pv}{\partial_{\nu}}
\newcommand{\au}{A_{\mu}}
\newcommand{\av}{A_{\nu}}
\newcommand{\p}{\partial}
\newcommand{\ts}{\times}
\newcommand{\ld}{\lambda}
\newcommand{\al}{\alpha}
\newcommand{\bt}{\beta}
\newcommand{\ga}{\gamma}
\newcommand{\si}{\sigma}
\newcommand{\ep}{\varepsilon}
\newcommand{\vp}{\varphi}
\newcommand{\zt}{\mathrm}
\newcommand{\bb}{\mathbf}
\newcommand{\dg}{\dagger}
\newcommand{\og}{\omega}
\newcommand{\Ld}{\Lambda}
\newcommand{\m}{\mathcal}
\newcommand{\dm}{{(k)}}

\title{Quantum optical metrology in the lossy $SU(2)$ and $SU(1,1)$ interferometers}
\author{Yang Gao}
\email{gaoyangchang@outlook.com}
\affiliation{Department of Physics,
Xinyang  Normal University, Xinyang, Henan 464000, People's Republic
of China}

\date{\today }

\begin{abstract}
%insert abstract here
We study the phase sensitivity in the conventional $SU(2)$ and
nonconventional $SU(1,1)$ interferometers with the coherent and
squeezed vacuum input state via the quantum Cramer-Rao bound. We
explicitly construct the detection scheme that gives the optimal
phase sensitivity. For practical purposes, we show that in the
presence of photon loss, both interferometers with proper homodyne
detections, are nearly optimal. We also find that unlike the
coherent state and squeezed vacuum state, the effects of the
imperfect detector on the phase sensitivity cannot be asymptotically
removed for a generic coherent-squeezed state by increasing the
amplifier gain of the OPA before the final detection.
\end{abstract}

%\keywords{}
\maketitle

\section{Introduction}

Recently, the first direct detection of a gravitational wave has
been carried out \cite{gw}. The prototype of such a detection is to
measure a relative phase shift between the two arms of an
interferometer due to passing a gravitational wave. The classical
setup is to feed the interferometer with a coherent and vacuum
state. The phase sensitivity of this strategy is bounded by the
shot-noise limit (SNL) \cite{caves}. The SNL {\it per se} is
remarkable enough to accomplish the sensitive detection of a
gravitational wave. However, further improvements are still needed
for more events.

One task of quantum optical metrology is to find the ultimate limits
on the phase sensitivity and the states that achieve these limits
\cite{qq}. In the absence of photon loss, employing quantum
resources such as $N00N$ states and entangled states \cite{noon}, it
is possible to improve the phase sensitivity from the SNL to the
Heisenberg limit (HL) \cite{hl}. While the nonclassical states give
enhanced phase sensitivities, their generations are known to be
highly difficult and resource intensive. Moreover, these states with
number of photons are very fragile to photon losses. A more
practical strategy is using a coherent and squeezed vacuum as the
input, which gives better phase sensitivity than the SNL
\cite{caves,css,opt}. For an ideal interferometer, the optimality of
this strategy is proved in Ref. \cite{opt}. Another possibility to
beat the SNL is to use the nonconventional interferometer first
proposed in Ref. \cite{su}, such as the optical-parametric amplifier
(OPA)-based interferometer \cite{su11}.

\begin{figure}[t!]
\centering \subfigure[]{
\includegraphics[width=0.5\columnwidth]{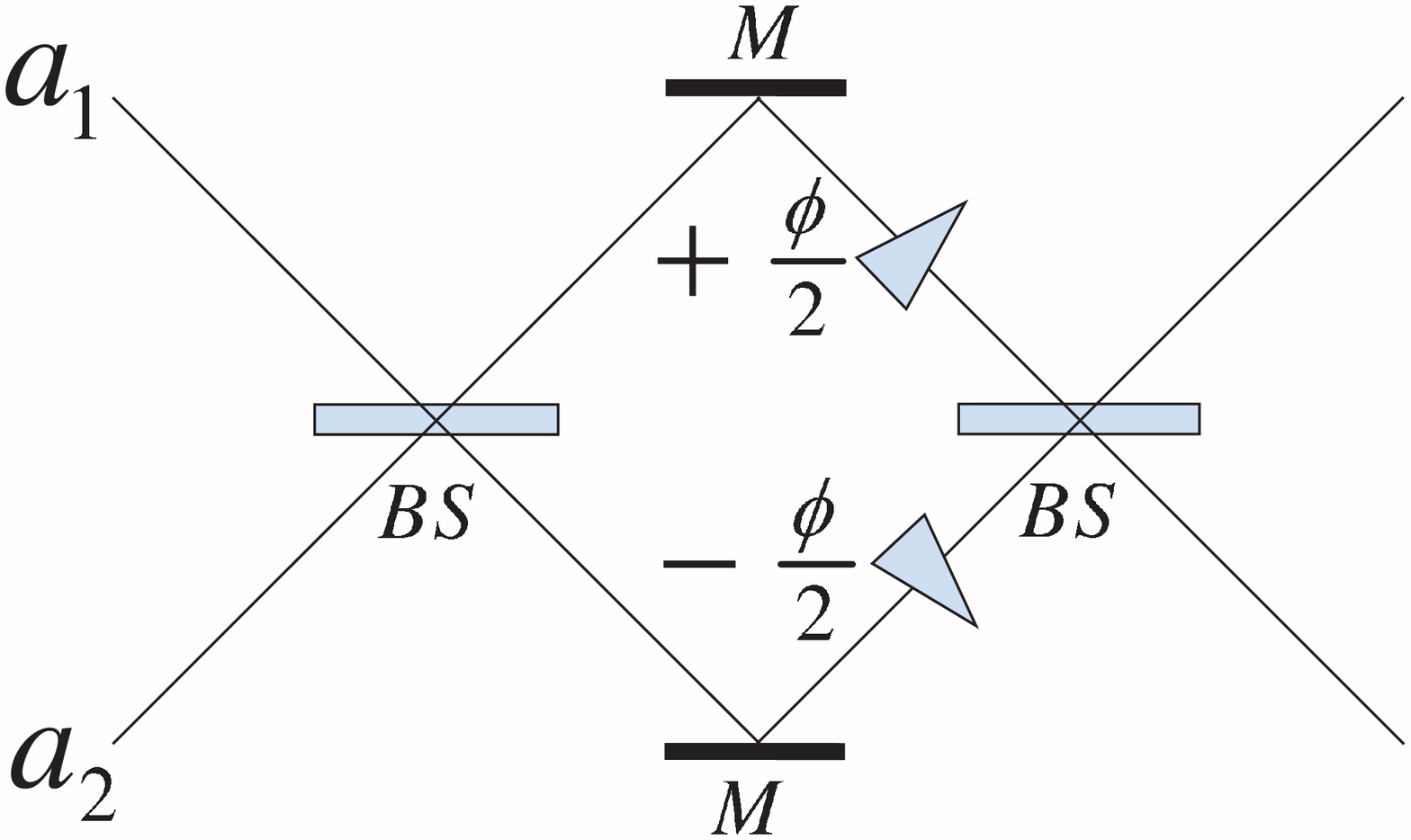}}\hspace{0.01in}
\subfigure[]{
\includegraphics[width=0.7\columnwidth]{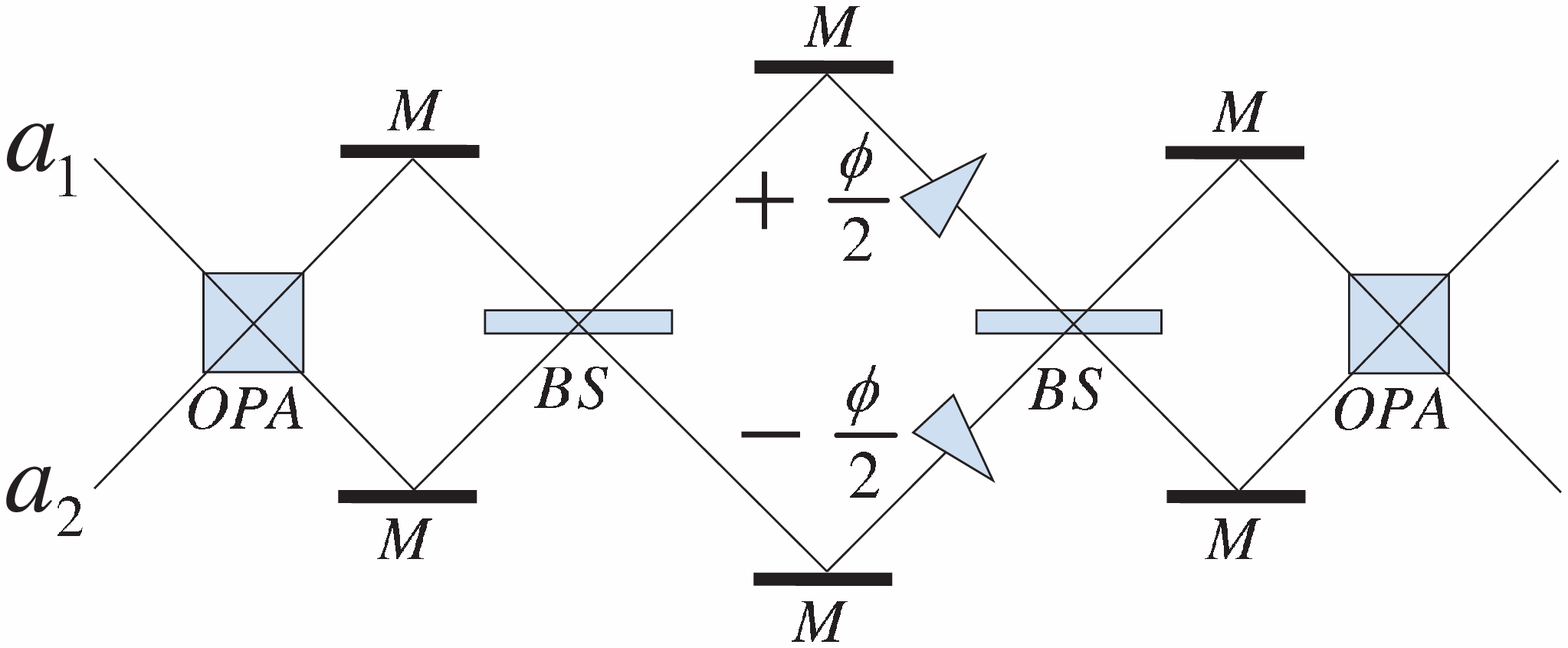}}\hspace{0.01in}
\caption{The schematics of (a) $SU(2)$ and (b) $SU(1,1)$
interferometers. There is a $\pi$ phase difference between the two
OPAs. $\phi$: relative phase shift between the two arms; $M$:
mirror.}
\end{figure}

In this paper, we will use the Gaussian input states to examine
their optimal performances in the conventional $SU(2)$ and
nonconventional $SU(1,1)$ interferometers. We find that the
detection schemes that gives the optimal phase sensitivities can be
explicitly constructed based on the quantum Cramer-Rao bounds (QCRB)
\cite{uncert,qcrb}, which can be realized by the balanced homodyne
detection associated with intensity-intensity correlations
\cite{detect}. The effects of internal and external photon losses on
the phase sensitivity will be investigated. In the presence of the
internal loss, we will illustrate that both interferometers with
proper homodyne detections, are nearly optimal for practical
situations. { Especially, the previously known results for the
equally lossy $SU(2)$ interferometer are generalized to the
unequally lossy case. We will find that the squeezed component of
the input state in the lossy $SU(1,1)$ interferometer is effectively
unnecessary. It will be also shown that by increasing the amplifier
gain of the OPA before the final detection, the effects of the
detector inefficiency on the phase sensitivity cannot be
asymptotically removed for a generic coherent-squeezed state, except
for the coherent state \cite{exloss} and the squeezed vacuum state
\cite{paris2}.}

The organization of this paper is as follows. In Section II we
review the QCRB with the Gaussian states, and construct the relevant
operators (describing the detection scheme) that could hit the QCRB.
Some examples for the phase sensitivity in the $SU(2)$
interferometer are discussed in Section III. Section IV contains
some examples in the $SU(1,1)$ interferometer. Then we consider the
effects of the detection inefficiency on the QCRB in Section V.
Finally, we end with a short conclusion.

\section{QCRB with Gaussian states and optimal detections}

We suppose that the phase shift $\phi$ is encoded in the output
state $\rho$ of the photons, after passing the interferometer. The
QCRB sets the ultimate limit for the phase sensitivity $\Delta \phi$
in the quantum optical metrology \cite{qq,uncert}. It is related to
the quantum Fisher information (QFI) $F$ by \be \Delta \phi \ge
\frac{1}{\sqrt{F}}. \ee The QFI is given by the variance of the
so-called symmetric logarithmic derivative (SLD) $L$, \be F
&=&\Delta ^2 L =\me {L^2}, \\ \rho' &=& \frac{1}{2}(\rho L+L\rho),
\ee where the notations $\Delta^2 O=\me{ (O-\me O)^2}$ in terms of
$\me O=\zt {Tr}[\rho O]$, and $O'=\p_\phi O$ for an arbitrary
operator. Here the fact $\me L=0$ has been used. The analytical
expression or numerical calculation of the QFI usually pose
formidable challenges. One exception is for the Gaussian state
\cite{qcrb}, which is completely characterized by the mean vector
${\bb v}$ and covariance matrix ${\bb \Sigma}$. For two-mode
Gaussian state with annihilation operators $a_1$ and $a_2$, we have
\be {\bb v} =\me {\bb a }, \quad {\bb \Sigma}= \me {(\tilde{\bb a}
\cdot\tilde{\bb a}\!\!\!\ ^{\top})}, \ee where ${\bb a}\!\!\!\
^{\top}=(a_1, a_1^\dg, a_2, a_2^\dg)$, $\tilde{{\bb a}}={\bb a}-{\bb
v}$, and $(\cdot)$ means the symmetrized ordering product. The SLD
and QFI for the Gaussian state in the matrix forms are given by
\cite{qcrb}, \be L &=& \frac{1}{2}\tilde{{\bb a}}\!\!\!\ ^{\top}
{\bb A} \tilde{\bb a}-\frac{1}{2}\zt{Tr}[\bb {\Sigma A}] +
\tilde{{\bb a}}\!\!\!\ ^{\top} {\bb b} \nn F &=& \frac{1}{2}\zt {Tr}
[{\bb \Sigma}'{\bb A}]+{\bb v}'\!\!\!\ ^{\top} {\bb b }, \label{5}
\ee  { where the matrix elements are \be {\bb
A}_{jk}=\frac{1}{2}[(\mathbf{\Sigma}\otimes\mathbf{\Sigma}
+\mathbf{\Omega}\otimes\mathbf{\Omega}/4)^{-1}]_{jk,pq}{\bb
\Sigma}'_{pq} \ee in terms of $\bb \Omega_{jk}= [\bb a_j,\bb a_k]$,
and the vector is ${\bb b}=\mathbf{\Sigma}^{-1}\mathbf{v}'$. Here
the direct-product of two arbitrary matrices $A$ and $B$ is defined
by $(A \otimes B)_{jk,pq}=A_{jp}B_{kq}$.}

Since the QCRB for the phase sensitivity is asymptotically
attainable for a large number of repeated measurements, it is
imperative to specify a detection scheme that could achieve the
QCRB. For the Gaussian state, we take the detection scheme as \be
M=\frac{1}{2}\tilde{{\bb a}}\!\!\!\ ^{\top} {\bb A}_0 \tilde{\bb a}
+ {\bb a}\!\!\!\ ^{\top} {\bb b}_0, \ee where the matrix $\bb A_0$
and the vector $\bb b_0$ can be freely chosen for our purpose. The
detection signal is thus given by $\me M=\zt {Tr}[\bb {\Sigma
A}_0]/2+{\bb v}\!\!\!\ ^{\top} {\bb b}_0$. The rate of this signal
change as a function of phase is \be \p_\phi \me M = \frac{1}{2}\zt
{Tr}[\bb \Sigma' \bb A_0]+\bb v'\!\!\!\ ^{\top} {\bb b}_0. \label{7}
\ee Meanwhile, the detection noise is described by $\Delta^2 M=\me{
(M-\me M)^2}$, where \be M-\me M= \frac{1}{2}\tilde{{\bb a}}\!\!\!\
^{\top} {\bb A}_0 \tilde{\bb a} - \frac{1}{2}\zt {Tr}[\bb {\Sigma
A}_0]+\tilde {\bb a}\!\!\!\ ^{\top} {\bb b}_0 . \label{8} \ee The
phase sensitivity is then estimated as the ration of the detection
noise versus the rate of the signal change, \be \Delta^2
\phi=\frac{\Delta^2 M}{|\p_\phi \me M|^2}. \label{9} \ee Now
assuming $\bb A_0=\bb A$ and $\bb b_0=\bb b$, from Eqs.
(\ref{5},\ref{7},\ref{8}) we find that $M-\me M$ and $\p_\phi \me M$
become the SLD and QFI, respectively. So Eq. (\ref{9}) takes \be
\Delta^2 \phi=\frac{\me {L^2}}{F^2}=\frac{1}{F} \ee for $F=\me
{L^2}$. Therefore, the attainability of the QCRB for the Gaussian
state is constructively proved.

Compared with the previously known detection schemes such as parity,
intensity, intensity difference, homodyne detections and etc., the
$M$-detection could always show the best performance among all
possible detections irrespective of photon loss. Though the parity
detection gives the optimal sensitivity under the lossless
conditions, it suffers greatly under the lossy conditions. Because
the quadratic form of the $M$-detection, its physical realization
can be simply implemented by the balanced homodyne detection
associated with intensity-intensity correlations according to Ref.
\cite{detect}. Furthermore, the explicit form of $M$-detection
allows us to assess the performances of other schemes, and shows the
way to better detection schemes.

\section{Phase sensitivity in the $SU(2)$ interferometer}

\begin{figure}[t!]
\centering \subfigure[]{
\includegraphics[width=0.45\columnwidth]{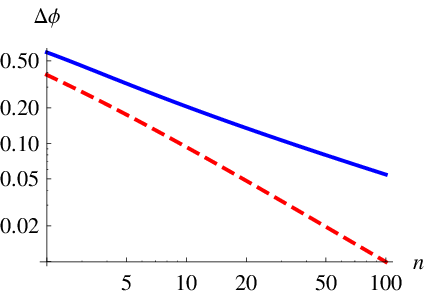}}\hspace{0.01in}
\subfigure[]{
\includegraphics[width=0.45\columnwidth]{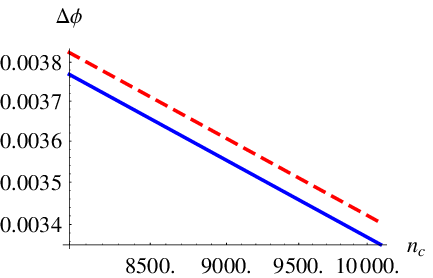}}\hspace{0.01in}
\caption{ The phase sensitivity in the MZI with a coherent and
squeezed vacuum input state. (a) The optimized phase sensitivity
over $n_s$ as a function of the total average photon number
$n=n_c+n_s$. Dashed: ideal MZI ($\xi=1$); Solid: lossy MZI
($\xi=0.8$). (b) The phase sensitivity with the homodyne detection
as a function of the average coherent photon number $n_c$. Dashed:
$p_2$-detection; Solid: $M$-detection. Here $\xi_1=0.8$ and
$\xi_2=1$.}
\end{figure}

In the traditional Mach-Zehnder interferometer (MZI) \cite{su} as
shown in Fig. 1(a), the output signal is very sensitive to the
relative phase shift between the two arms. The ability to resolve
extremely small relative phase shift finds many applications in
optical gyroscopes, gravitational wave detection, quantum imaging
and sensing \cite{dow}. It is more convenient to use the $SU(2)$
group formalism for this passive lossless four-port optical system.
The propagation of photons through the MZI can be described by the
composition of three $SU(2)$ transformations $T_\zt{MZI}=T_\zt{BS}
T_\phi T_\zt{BS}$: $\bb a \to T_\zt{MZI} \bb a$, where \be
T_\zt{BS} &=& \frac{1}{\sqrt{2}}\lz \bx 1 & 0 & 1 & 0 \\
0 & 1 & 0 & 1 \\ 1 & 0 & -1 & 0 \\ 0 & 1 & 0 & -1 \ex \rz, \nn
T_\phi &=& \lz \bx e^{{i\phi\over2}} & 0 & 0 & 0 \\
0 & e^{-{i\phi\over2}} & 0 & 0 \\ 0 & 0 & e^{-{i\phi\over2}} & 0 \\
0 & 0 & 0 & e^{{i\phi\over2}}\ex \rz, \ee where $T_\zt{BS}$ and
$T_\phi$ represent the actions of the beam splitter and the
symmetric phase shift \cite{rela}, respectively. Without the phase
shift, $T_\zt{BS}T_\zt{BS}=1$ and the output is equal to the input.

With a coherent state $\lg \al_1 \lg 0_2 =e^{\al(a_1^\dg-a_1)}\lg
0_1 \lg 0_2$ into the MZI, the mean vector and covariance matrix of
the input state are \be \bb v_{in}^\top &=& (\al,\al,0,0), \nn
\bb \Sigma_{in} &=& \frac{1}{2}\lz \bx 0 & 1 & 0 & 0 \\
1 & 0 & 0 & 0 \\ 0 & 0 & 0 & 1 \\ 0 & 0 & 1 & 0 \ex \rz. \ee After
the transformation $ T_\zt{MZI}$, the output state is represented by
\be \bb v_{out} &=& T_\zt{MZI} \bb v_{in} = \al \lx
\cos\frac{\phi}{2},\cos\frac{\phi}{2},i\sin\frac{\phi}{2},-i\sin\frac{\phi}{2}\rx^\top,
\nn \bb \Sigma_{out}  &=&  T_\zt{MZI}\bb \Sigma_{in}
T_\zt{MZI}^\top=\bb \Sigma_{in}. \ee For simplicity, we assume that
the estimated phase is around $\phi=0$ from now on. The calculations
lead to the QFI $F=\al^2$, and the $M$-detection $M \propto p_2=i
(a_2^\dg-a_2)/\sqrt{2}$, namely, a homodyne detection. The optimal
phase sensitivity is thus the SNL, $\Delta \phi = 1/\sqrt{n_c}$,
where $n_c=\al^2$ is the average photon number in the coherent beam.

It has been shown that the SNL can be beaten by taking advantage of
quantum resources towards achieving the more fundamental HL
\cite{hl}, i.e., $\Delta \phi \simeq 1/n$, where $n$ is the average
photon number inside the MZI. Considering the practical difficulty
in creating the required quantum resource (such as creating high
$N00N$ states, large-gain squeezed light sources), we take the
moderate strategy that using a coherent and squeezed vacuum state
$\lg \al_1 \lg {0,r}_2 =e^{\al(a_1^\dg-a_1)}e^{r(a_2^{\dg
2}-a_2^2)/2}\lg 0_1\lg 0_2$ as the input \cite{caves,css,opt}. It is
described by \be \bb v_{in}^\top &=& (\al,\al,0,0),
\nn \bb \Sigma_{in} &=& \frac{1}{2}\lz \bx 0 & 1 & 0 & 0 \\
1 & 0 & 0 & 0 \\ 0 & 0 & Y & X \\ 0 & 0 & X & Y \ex \rz, \ee where
$X=\cosh 2r$, $Y=\sinh 2 r$, and $r$ is the squeezing factor. Then
output state is represented by { $\bb v_{out} = T_\zt{MZI} \bb
v_{in}$ and $ \bb \Sigma_{out} = T_\zt{MZI}\bb \Sigma_{in}
T_\zt{MZI}^\top$}. The corresponding QFI is given by $F=\al^2
e^{2r}+\sinh^2 r$, and the $M$-detection is \be M &=& i \sinh r
[(\tilde{a}_1\tilde{a}_2^\dg-\tilde{a}_1^\dg \tilde{a}_2)\sinh
r+(\tilde{a}_1^\dg \tilde{a}_2^\dg-\tilde{a}_1\tilde{a}_2)\cosh
r]\nn && +i \al (1+2e^r\sinh r) (a_2^\dg-a_2). \ee It can be seen
that for a given average photon number $n=n_c+n_s=\al^2+\sinh^2 r$,
the phase sensitivity reaches the HL at the optimal point $n_c
\simeq n_s$. However, due to technical limitations, $n_s \ll n_c$
and $n \approx n_c$, we actually have $\Delta \phi \simeq
e^{-r}/\sqrt{n}$, which can still beat the SNL for a positive
squeezing factor. Moreover, the $M$-detection is approximately a
homodyne detection $M \propto p_2$.

Next, we consider photon loss to the environment inside the
interferometer. It can be modeled by a fictitious beam splitter at
the middle of the interferometer arm, which transforms the beam as
follows: $ a_k \to \mathcal{L} a_k = a_k\sqrt{\xi_k} +\upsilon_k
\sqrt{1-\xi_k} $ ($k=1,2$), where $\xi_k$ is the photon
transmissivity, and $\upsilon_k$ is the environment mode in the
vacuum state. The actions of $\m L$ on the mean vector and
covariance matrix take the form of \be \bb v \to R \bb v, \quad \bb
\Sigma \to R \bb \Sigma R^\top+\bb
\Sigma_{\mathrm{vac}}, \ee where \be R &=& \lz \bx \xi_1 & 0 & 0 & 0 \\
0 & \xi_1 & 0 & 0 \\ 0 & 0 & \xi_2 & 0 \\ 0 & 0 & 0 & \xi_2 \ex \rz,
\nn \bb \Sigma_\mathrm{vac} &=& \frac{1}{2} \lz \bx 0 & 1-\xi_1 & 0 & 0 \\
1-\xi_1 & 0 & 0 & 0 \\ 0 & 0 & 0 & 1-\xi_2 \\ 0 & 0 & 1-\xi_2 & 0
\ex \rz. \ee { Using this formalism, the QCRB for a coherent and
squeezed vacuum state into the MZI can be obtained. For a coherent
state ($r=0$), it gives the phase sensitivity $\Delta \phi =
1/\sqrt{\xi n_c}$, which is a constant factor $\xi=(\xi_1+\xi_2)/2$
reduction of the ideal case. To investigate the scaling of phase
sensitivity versus the average photon number $n$, we take the
optimization over $n_s$ under the constrain $n=n_c+n_s$. Using the
large-$n$ expansion, we find that \be \Delta^2 \phi = \lz \frac{\xi
n}{1-\xi}+O(\sqrt{n}) \rz^{-1}, \label{loss2} \ee and the optimal
value of $n_s$ is $n_s = O(\sqrt{n}) \ll n$.} We can see from Fig.
2(a) that in the presence of photon loss, the phase sensitivity
approaches the SNL for large $n$. As a comparison, for the lossless
case, we note that $n_s \simeq n/2$ and $\Delta \phi \simeq 1/n$.

For the practical cases, $n_s \ll n_c$ and $n \approx n_c$, we have
the approximate QCRB, \be \Delta^2 \phi \simeq \lz \frac{\xi
n}{1-\xi+\xi e^{-2r}}\rz^{-1}. \label{loss1} \ee If the squeezing
factor $r$ is sufficiently large, i.e., $e^{2r} \gg \xi/(1-\xi)$, we
further obtain $\Delta^2 \phi \simeq (1-\xi)/(\xi n)$, which is just
the optimal phase sensitivity in Eq. (\ref{loss2}). As for the
$M$-detection scheme, it effectively becomes a generalized homodyne
detection, \be M \propto p_2+{\sqrt{\xi_1}-\sqrt{\xi_2} \over
\sqrt{\xi_1}+\sqrt{\xi_2}}p_1, \label{gho} \ee which reduces to
$p_2$-detection for the balanced lossy arms $\xi_1=\xi_2$
\cite{su1cs}. With this $M$-detection, the phase sensitivity yields
Eq. (\ref{loss1}). On the other hand, the phase sensitivity based on
the $p_2$-detection is \be \Delta^2\phi \simeq \lz\frac{\zeta n
}{1-\zeta+\zeta e^{-2r}}\rz^{-1}, \ee where
$\zeta=(\sqrt{\xi_1}+\sqrt{\xi_2})^2/4$. As shown in Fig. 2(b), for
the unbalance lossy arms, the generalized homodyne detection gives
better result than the simple $p_2$-detection and approaches Eq.
(\ref{loss2}) as $n_c \to \infty$. { That is to say, in the presence
of photon loss, the MZI with the practical coherent and squeezed
vacuum state, plus the $M$-detection, performs nearly optimal. This
generalizes the previously known results for the equal lossy arms in
Refs. \cite{fql}.}

\section{Phase sensitivity in the $SU(1,1)$ interferometer}

To beat the SNL, Yurke {\it et al.} proposed a new type of
interferometer in which the beam splitters of the traditional MZI is
replaced with active four-port wave mixers \cite{su}. Unlike the
traditional MZI described by the $SU(2)$ group, this type of
interferometer is described by the $SU(1,1)$ group. For the setup in
Refs. \cite{su11}, two OPAs replace two beam splitters in the
traditional MZI. { In Fig. 1(b), we slightly modify the setup in
Refs. \cite{su11} by placing the traditional MZI between two OPAs in
order to avoid the phase ambiguity discussed in Ref. \cite{rela}.}
The transformation between the input and output modes is then given
by $\bb a \to T^-_\zt{OPA} T_\zt{MZI} T^+_\zt{OPA} \bb
a$, where \be T^\pm_\zt{OPA} &=& \lx \bx \mu & 0 & 0 & \pm \nu \\
0 & \mu & \pm \nu & 0 \\ 0 & \pm \nu & \mu & 0 \\ \pm \nu & 0 & 0 &
\mu \ex \rx, \ee where $T^{\pm}_\zt{OPA}$ represent the actions of
the OPAs. Here $\mu=\cosh g$, $\nu=\sinh g$, and $g$ is the
parametrical strength in the OPA process. Note that
$T^-_\zt{OPA}T^+_\zt{OPA}=1$. Thus in the absence of phase shift,
the input equals the output.

We first consider the ideal $SU(1,1)$ interferometer. The QFI with
the coherent input state $\lg \al_1 \lg 0_2 $ is calculated as \be F
= G(G+2)(2n_c+1)+n_c, \label{su11} \ee where $G=2\sinh^2 g$ is the
spontaneous photon number emitted from the first OPA. We notice that
Eq. (\ref{su11}) is different from the corresponding result in Ref.
\cite{su1cs}, because we used the symmetric phase shift between the
two arms instead of the biased phase shift. Without the OPA process
($G=0$), $F=n_c$ gives the SNL for a coherent state. For the vacuum
input ($n_c=0$), the state after the first OPA is the two-mode
squeezed vacuum, and the QFI becomes $F=G(G+2)$. Because the average
photon number inside the interferometer is just $n=G$, we have
$\Delta\phi=1/\sqrt{n(n+2)}$ \cite{tmsv}. However, due to technical
limitations, $n_c \gg G$, the actual phase sensitivity becomes
$\Delta^2\phi \simeq 1/(2n_c G^2)$ when the amplification of the
input signal gets large, i.e., $G\gg 1$. So the phase sensitivity is
improved by a factor of $2G^2$ compared with the SNL.

If we use $\lg \al_1 \lg {0,r}_2 $ as the input state, the QFI is
calculated as \be F &=& G (G+2) \left[n_c+\left(2
n_s+1\right)\left(n_s+1\right) \right]\nn && +(G+1)^2 \left(n_c e^{2
r}+n_s\right). \label{ideal} \ee We note that the average photon
number inside the interferometer is given by $n=(G+1)(n_c+n_s)+G$.
For the practical cases, $n_c \gg \max\{G,n_s\} \gg 1$, the QCRB
becomes \be \Delta^2\phi \simeq {1 \over 4 n_c n_s G^2}. \ee That
is, the phase sensitivity is further enhanced by a factor of $2n_s$
with respect to the coherent input state. For the MZI with the same
input state, the relevant phase sensitivity is $\Delta^2\phi \simeq
{1/(4n_c n_s) }$. Therefore, the $SU(1,1)$ interferometer performs
better than the MZI by a factor of $G^2$ (see Fig. 3(a)).

Now we study the effects of photon loss on the $SU(1,1)$
interferometer. Following the similar steps as the MZI case, the
QCRB for the lossy interferometer can be calculated. The numerical
results are plotted in Fig. 3(b). It can be seen that for
$\max\{G,n_s\} \gg 1$, and $n_c \to \infty$, we have \be
\Delta^2\phi \to {1-\xi \over \xi n_c} \ee for the MZI with the
input state $\lg \al_1 \lg {0,r}_2$, and \be \Delta^2\phi \to
{1-\xi_1 \over \xi_1 n_c G } \ee for the $SU(1,1)$ interferometer
with the input state $\lg \al_1 \lg {0}_2$ or $\lg \al_1 \lg
{0,r}_2$. For the balanced lossy arms ($\xi_1=\xi_2=\xi$), both of
them approach $\Delta^2\phi \to {(1-\xi)/(\xi n)}$, i.e., the
fundamental lower bound of the phase sensitivity derived in Refs.
\cite{fql}. This implies that replacing the second vacuum port with
a squeezed vacuum is almost unnecessary for a lossy $SU(1,1)$
interferometer. Finally, the optimal $M$-detection is also
approximately a homodyne detection.

\begin{figure}[t!]
\centering \subfigure[]{
\includegraphics[width=0.45\columnwidth]{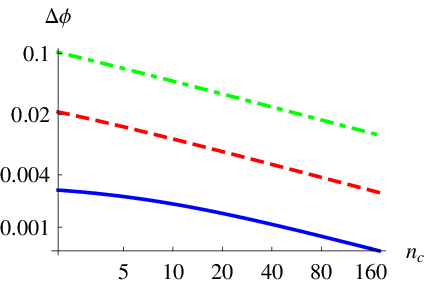}}\hspace{0.01in}
\subfigure[]{
\includegraphics[width=0.45\columnwidth]{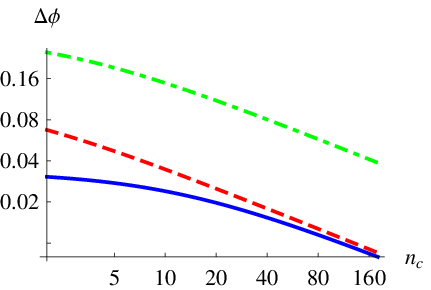}}\hspace{0.01in}
\caption{ Loglog-plots of the phase sensitivity $\Delta \phi$ in the
$SU(1,1)$ interferometer as a function of the average coherent
photon number, $n_c=\al^2$. Solid: coherent and squeezed vacuum
state; dashed: coherent state. As a comparison, the dot-dashed lines
represent the results with a coherent and vacuum state into the MZI.
The common parameters $G=20$, $n_s=10$. (a) $\xi=1$; (b) $\xi=0.8$.}
\end{figure}

\section{Effects of external loss}

In the previous sections we only consider the effects of photon loss
to the environment insider the interferometer. Now we investigate
the photon loss at the detectors, due to imperfect detectors
\cite{exloss,paris2}. This external loss can also be modeled by a
fictitious beam splitter as the internal loss, where the
transmissivity parameter $\xi_k$ is related to the detection
inefficiency. { Ignoring the internal photon loss and assuming
$\xi_1=\xi_2=\xi$, the QFI for the input state, $\lg \al_1 \lg
{0,r}_2$, in the $SU(1,1)$ interferometer becomes \be F &=& {\xi}
\bigg\{\frac{G (G+2)}{2} \left[2n_c+{\xi} +\frac{ (2n_s+1)^2\xi}{
2n_s\xi(1-\xi)+1}\right]\nn && +(G+1)^2 \left(\frac{n_c}{1-\xi +\xi
e^{-2r}}+n_s\right)\bigg\}. \ee For the ideal detector, $\xi=1$, we
get Eq. (\ref{ideal}). For a vacuum input, $\al=r=0$, $F=\xi^2
G(G+2)$, which reduces the ideal result by a factor of $\xi^2$. For
the practical situations $n_c \gg \max\{G, n_s\}\gg 1$, $F \simeq
\xi (2-\xi) n_c G^2/ (1-\xi) $, which also means that the squeezed
component of the input state is almost unnecessary. If the second
OPA is not introduced, namely, $T^-_\zt{OPA}=1$, the corresponding
QCRB is plotted in Fig. 4(a) as a comparison. It indicates that the
implement of the second OPA improves the phase sensitivity over that
without the OPA before the final detection.}

\begin{figure}[t!]
\centering \subfigure[]{
\includegraphics[width=0.45\columnwidth]{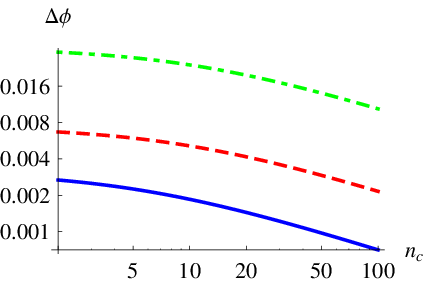}}\hspace{0.01in}
\subfigure[]{
\includegraphics[width=0.45
 \columnwidth]{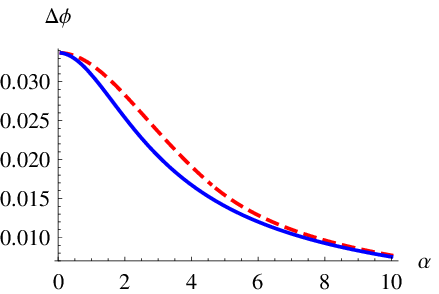}}\hspace{0.01in}
\caption{ The plots of QCRB in the presence of external photon loss.
(a) The QCRB in the $SU(1,1)$ interferometer with $G=20$.
Dot-dashed: without the second OPA; Dashed: with the second OPA;
Solid: ideal detector. (b) The QFI for a single-mode
coherent-squeezed state with a large amplifier gain $G\gg 1$. The
common parameters are $n_s=20$ and $\xi=0.8$.}
\end{figure}

{ Next, we consider the following question: whether the effects of
the detector inefficiency on the phase sensitivity can be
effectively removed by an OPA (with gain $G$) before taking the
final detection. For simplicity, a single-mode analog of the
interferometer is invoked. We take a coherent-squeezed state, $\lg
{\al, r}=e^{\al(a^\dg -a)}e^{r(a^{\dg 2}-a^2)/2}\lg 0$, as the input
to the phase shifter $T_\phi=e^{-i \phi a^\dg a}$. Then an OPA is
introduced by $T_\zt{OPA}^-=e^{g(a^2-a^{\dg 2})/2}$. At last, the
external loss described by $\m L a=
a\sqrt{\xi}+\upsilon\sqrt{1-\xi}$ is applied. The combined
transformation is thus given by \be T_\zt{tot} = \m L T_\zt{OPA}^-
T_\phi \label{tot} \ee Our task is to optimize the corresponding
QCRB over the adjustable parameters $\phi, G$. We find that for a
given $\phi$, the relevant QFI is a increasing function of $G$. As
$G \to \infty$, we obtain \be F \to  4\cos^2\phi {\al^2 X + Y^2 - Y
(\al^2 + Y) \cos 2\phi  \over (X - Y \cos 2\phi)^2}. \ee The
optimization of this QFI over $\phi$ is plotted in Fig. 4(b). We see
that the QCRB in the presence of the external photon loss is always
smaller than the ideal case. Only for some exceptional points,
$\al=0$ (squeezed vacuum state) and $r=0$ (coherent state), they can
be equal to $\Delta \phi = 1/(2\al)$ and $\sqrt{2}Y$, respectively
\cite{exloss,paris2}. In other words, by increasing the amplifier
gain, the effects of the detector inefficiency on the QCRB cannot be
asymptotically canceled out for a generic coherent-squeezed state.
Similar results can be obtained for the two-mode case.}

\section{Conclusion}

In summary, we have used the coherent and squeezed vacuum input
state to study its optimal performances in the traditional MZI and
the OPA-based MZI via the QCRB. Based on the SLD, we have explicitly
constructed the detection schemes that gives the optimal phase
sensitivities. Considering the technical limitations, we have shown
that in the presence of photon loss, both of the interferometers
with a proper homodyne detection are nearly optimal.  { Especially,
we have generalized the previously known results for the equally
lossy $SU(2)$ interferometer to those for the unequally lossy case.
We have also found that the squeezed component of the input state in
the lossy $SU(1,1)$ interferometer is effectively unnecessary.
Finally, we have found that the effects of the detector inefficiency
on the QCRB cannot be asymptotically canceled out for a generic
coherent-squeezed state by increasing the amplifier gain of the
OPA.}

\begin{acknowledgments}
The author would like to acknowledge the support from NSFC Grand
No. 11304265 and the Education Department of Henan Province (No.
12B140013).
\end{acknowledgments}

%Create the reference section using BibTeX:

\end{document}